# Understanding students' difficulties in terms of coupled epistemological and affective dynamics

Ayush Gupta, Brian A. Danielak, Andrew Elby
University of Maryland, College Park, ayush@umd.edu, briandk@umd.edu, elby@umd.edu

*Abstract* - **An established body of literature shows that a student's affect can be linked to her epistemological stance [1]. In this literature, the epistemology is generally taken as a belief or stance toward a discipline, and the affective stance applies broadly to a discipline or classroom culture. A second, emerging line of research, however, shows that a student in a given discipline can shift between multiple locally coherent epistemological stances [2]. To begin uniting these two bodies of literature, toward the long-term goal of incorporating affect into fine-grained models of in-the-moment cognitive dynamics, we present a case study of "Judy", an undergraduate engineering major. We argue that a fine-grained aspect of Judy's affect, her annoyance at a particular kind of homework problem, stabilizes a context-dependent epistemological stance she displays, about an unbridgeable gulf she perceives to exist between real and ideal circuits.**

## INTRODUCTION

As instructors, educators, and education researchers, intuitively we know that students' emotions, motivations, and sense of identity can strongly influence their behaviors in science learning environments. Research in cognitive science makes a persuasive case for the role that affect and identity play in regulating performance on cognitive tasks [3-7]. Nonetheless, explicitly incorporating emotion and identity into theories of science learning, though recognized as important, has proved to be a challenge [8-11]. Cognitivist accounts of student reasoning and conceptual change in science [12-16] often pay little attention to the role of affect and identity in the use of knowledge by learners. On the other hand, accounts of learning and conceptual change that have paid greater attention to learners' affect, social interaction, and the development of identity associated with the learning process [9],[17-20] often treat those subjects at a coarse grain size. They typically attend to students' stances towards a subject as a whole or their aggregate emotions towards particular styles of learning. For example, a student's negative emotional reaction towards a traditional, authoritarian classroom environment, and her sense of identity as someone who likes multiple interpretations can distance her from the subject of mathematics itself [17]. It is intuitively appealing, however, that students' emotional states and aspects of their emerging disciplinary identity play a role not only in shaping their choices and beliefs, but in stabilizing or destabilizing their reasoning *on a moment-to-moment basis*. In this paper, using a case study of an undergraduate engineering student, we aim to support the intuitively plausible fine-grained connection between affective and epistemological stances.

In two independent lines of research, researchers have started to build fine-grained models of learners' affect and personal epistemologies. Hammer and Elby [21] extended the ideas of society of mind [22] and fine-grained elements of conceptual reasoning [12] to fine-grained epistemological elements that activate in different combinations to constitute an individual's stance towards the nature and form of knowledge they are engaging with in a given moment. These stances are contextually activated, though coherent 'belief-like' patterns of association can emerge over time [21].

Education research, especially in mathematics, has also started to pay attention to affect by describing its main components as beliefs, attitudes, emotions, [8] and values [23],[24]. Researchers aiming at fine-grained models of learners' affect have approached it from a variety of perspectives: socio-cultural [25], positioning analysis [26] and as a cognitive representational system [27].

These fine-grained models of affect and epistemology, while not incompatible with one another, have remained separate. As such, connections between affect and epistemology have been limited to explorations at coarse-grained scales, as discussed above [1],[9],[17]. The need to integrate the emotional, epistemological, and conceptual aspects of cognition at a fine-grained-level is recognized as important but under-emphasized [4],[8],[11].

In this paper, we use a case study of an undergraduate engineering major to argue for the feasibility and explanatory power of integrating epistemological and affective elements into fine-grained models of cognitive dynamics. In some contexts, "Judy" argues that there is a wide gulf between real circuits and ideal circuits, a gulf that renders qualitative understanding of ideal circuits practically useless. We will argue that an affective stance, Judy's deep *annoyance* at qualitative, conceptual homework problems, stabilizes her epistemological stance. Disrupting the annoyance disrupts the epistemological stance.

This argument has instructional implications. If a learners' emotional stance plays a significant role in sustaining her epistemological stance, as we argue is the case with Judy, then targeting her epistemological stance directly [28] might not be as productive as addressing the affective stance that sustains the epistemological stance.

In the following section, we present the methods of our case study. We then present the patterns we found in Judy's epistemological and emotional stances. Finally, we make the case that a fine-grained account of Judy's reasoning is empirically inadequate without affective elements that lend stability to her epistemological stances. The argument relies on competing toy models of Judy's cognitive dynamics, one of which includes affect and the other of which does not. In closing, we discuss the instructional implications of our analysis.

## DATA AND METHODS

We videotaped clinical interviews of electrical engineering majors in a Basic Circuits course at a Large Public University. The course homework and exams mixed "traditional" quantitative problems with conceptual questions asking students to interpret equations and/or explain physical processes (rather than plugging and chugging). These conceptual problems were co-created by the course instructor and two of the authors (AG, AE). One of us (BD) conducted one-hour semi-structured interviews with Judy and with 3 other students to explore their approaches to mathematics within, and their views and feelings about, the course. Interview participants were paid ten dollars.

Working as a group, we looked for epistemological and affective patterns in participants' responses, both within and across subjects, borrowing tools from discourse analysis [29], framing analysis [30], and affective analysis [25] to interpret gestures, facial expressions, word choice, and the contextualized substance of their utterances. In this paper, we present our analysis of one of our participants, Judy – a second year Electrical Engineering and Physics major. We formed explanations for her reasoning and behaviors and then looked for evidence elsewhere in the data to confirm or disconfirm the validity of these patterns [31],[32]. Working from a knowledge-in-pieces perspective [12],[33], we did not expect global coherence in all aspects of her thinking; we continually considered how specific contextual cues might trigger different local coherences in her thinking [2],[12],[28].

## CASE STUDY: PATTERNS IN JUDY'S BEHAVIOR

In this section, we illustrate four patterns in Judy's reasoning and behavior about circuits:
1. Judy is annoyed by the conceptual questions
2. Judy views conceptual reasoning as unproductive to her learning
3. Judy sees a gulf between real-world circuits and the ideal circuits probed by the conceptual questions
4. When discussing equations, Judy displays no annoyance and does not discuss the real/ideal gulf from Pattern #3

At the first glance, it seems reasonable to explain Judy's behavior in terms of her epistemological stance (about the disconnect between real and ideal circuits) driving her view that conceptual reasoning about ideal circuits is unproductive. In such an account, her annoyance is a by-product that doesn't contribute additional explanatory power. Analysis of the entire interview, however, shows the greater explanatory power of a model in which Judy's *annoyance* plays a key role in mediating the connection between her epistemological stance about real vs. ideal circuits and her view that conceptual reasoning is unproductive

*(1) Annoyance at conceptual problems:*

Judy, like other students we interviewed, distinguishes between the regular, mathematical problems and the more conceptual problems on the homework:

> [15:33] Interviewer: A student whose opinion I heard earlier from your class had noticed that the homework and tests seemed to contain two sort of types of questions—two different kinds of questions. And I was wondering if it's been your experience that you've noticed something like that.

> [16:01] Judy: Oh. I think, um, you mean two parts, right? Umm, I think one of them is like, um, just problem-solving. Like, you have a diagram {*spreads hands out, gently taps table with fingertips*} and then you solve for the current or voltage. Um, and the other type is like {*shakes head gently side-to-side*} a physical question. They will ask you what is {*shakes head*} physically {*pinches thumb and forefingers of both hands together, moving her hands up and down*} happening in the circuit, and you have to explain them in {*shakes head slightly, left to right*} words.

Though our attempts to transcribe gestures cannot fully capture the annoyance Judy displays when discussing the "physical question[s]," we believe it was strongly present here and elsewhere in the interview. During one exchange, Judy explicitly labels these questions as "annoying":

> [17:13] Judy: Yeah. I mean those questions are—it's {*shakes head gently, but repeatedly*} kind of annoying. But it's good to know {*rise in voice tone*}.

Judy's annoyance is coupled to another pattern in her reasoning: seeing conceptual reasoning as unproductive.

*(2) Judy views conceptual reasoning as unproductive:*

Judy sees conceptual reasoning as useless for practical purposes and considers equation-based problem-solving as "more helpful" and "more important":

> [19:22] Interviewer: Do you think one kind of question is more helpful to you than the other kind?

[19:28] Judy: Umm, yeah I think the {*cocks head, smiles*} problem-solving is more helpful {*laughs*} and {*nods*} more important.

And, when the interviewer asks about the importance of "physical" questions to a professional engineer (as opposed to an engineering student), Judy feels those questions have little bearing on what engineers actually do:

[18:50] Judy: {*shakes head gently, but repeatedly, left-to-right*} those physical questions are *not very related to the actual world; not related* {*knits brow*} *to [a professional engineer's] job.* So...that's why I say *it's not very necessary* for student[s] who are *only* in EE major to learn those parts. (emphasis added)

*(3) Real/ideal gulf:*

Early in the interview, the interviewer asks Judy to imagine a female professional engineer who is taking a basic circuits course to reinforce her understanding of circuits, in preparation for a new job assignment. Specifically, the interviewer asks what a perfect version of the circuits course would look like for that engineer.

[05:16] Judy: Well, I think, um, the course should talk more about, um, how the actual world works. Because, sometimes *we talk about, like, ideal circuits and, um theoretical methods. Those are not related to the actual circuit and those kind of things.* So if the professor can talk a little bit more about the actual circuit and how those work, then it may be better for her. (emphasis added)

[05:55] Interviewer: Why is that difference important to you?

[05:57] Judy: Um, Because the ideal world is different.

Elsewhere, Judy and the interviewer clarify that the phrase "theoretical methods", to Judy, refers to conceptual, non-mathematical reasoning. Here and elsewhere, Judy expresses the view that the "real world" of circuits operates much differently from the "ideal circuits" she has learned about in class; the second is not a close approximation of the first. As a result, the class's conceptual focus on ideal circuits is a poor use of class time:

[06:13] Judy: Um, and, sometimes [in class] we talk about, um, the physical aspects of the circuits. Um, but I feel like *when you really work on circuits it's—it's not very important.* I mean, it's *better to know* what happened physically, but if those are not ideal circuits, then it's different. (emphasis added)

This unbridgeable gulf between ideal and real circuits, she says, is what makes the conceptual problems annoying; those problems call for students to use concepts that apply only to ideal circuits:

[08:53] Judy: Um, like a capacitor, I might say this wrong, but um, like they have a limit that how many currents, or how many electrons can stay on the plate. *That's a real world capacitor*. But, for, ideally, *sometimes in the course we assume that each plate can have infinite amount of electrons*, and those kind of things. So, I mean, *if that doesn't exist then why do we use that method to solve it? We are not gonna use that method in the future anyway.* (emphasis added)

Later in the interview, Judy says that she expects later, more advanced course will teach her new ways of thinking about circuits that apply to real circuits and that are totally different from what she is learning in her present class. Crucially, despite her strong views about the real/ideal gulf Judy's use of mathematics led us to our fourth observed pattern.

*(4) Non-activation of the real/ideal gulf during traditional problem-solving.*

Later in the interview, the interviewer shows Judy a simple DC circuit (figure 1) where the potential at a point a distance $\ell$ along the "fat resistive" wire is given by $V = (V_0 - \rho I \ell)$:

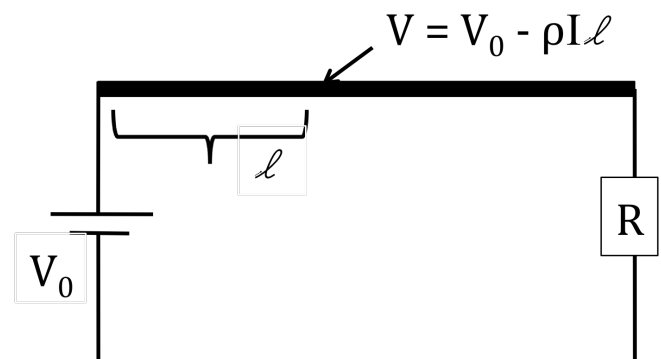

FIGURE 1
THE CIRCUIT FOR THE "FAT RESISTIVE WIRE" PROMPT

[27:28] Interviewer: The question is, um, how would you explain this equation to a friend from class?

[27:35] Judy: Um, first I would just, um, repeat this equation, and then explain what rho is, what I is, what L is. And then, um, I guess rho I and L is somewhat related to the resistance of the wire. So, I will explain that to him. And then, this is—yeah, I guess rho and L is kind of proportional to R. This question is basically just, uh, like uh, I equals V over R. Something like that. It's actually a lot more easier question than it looks like. {*smiles*} Yeah. (emphasis added)

…

[28:37] Interviewer: How did you know that um, rho and L were related to resistance?

[28:41] Judy: Um, because I learned that before {*smiles*}. And, it looks like, also looks like kind of proportional to the, um, V. Yeah, so. I have a feeling {*laughs*}.

When talking about or engaging in equation-based problem solving, Judy displays positive affect (she smiles and laughs and does not appear to be tensed), and we find no evidence that in these moments she is ever thinking about the real/ideal gulf (pattern#3). This is striking given that the equations she uses encode the same idealizing assumptions she disparages elsewhere in the interview. Earlier, though, it was Judy—not the interviewer—who brought up the real/ideal gulf when discussing conceptual questions. Her epistemological view about the real/ideal gulf is thus context-dependent: *present* when she discusses conceptual problems, but *absent* in both her discussions of and solutions to traditional ones.".

## MODELING JUDY'S REASONING AND BEHAVIOR: AFFECT MEDIATED EPISTEMOLOGICAL DYNAMICS

*Judy's epistemological stance about the real/ideal gulf is context dependent*

First, we argue that Judy's epistemological stance about the real/ideal gulf is indeed context dependent, i.e. that sometimes it is "on," playing an important role in her thinking, while at other times it is "off," allowing her to think in ways that are at odds with that stance. Pattern #2 and #3 suggest that, at least when thinking about the conceptual problems, Judy takes the stance that ideal circuits are so different from real ones as to render pointless thinking about ideal circuits. In pattern #4, by contrast, Judy does *not* exhibit this stance. We're *not* claiming that Judy takes an opposing epistemological stance when thinking about mathematical problem-solving. Rather, we're claiming that when she's focused on mathematical problem-solving the *issue* of a possible real/ideal gulf isn't part of her thinking. The four patterns discussed above are consistent with a context-dependent but affect-free toy model (figure 2) in which Judy's epistemological view about the gulf between real and ideal circuits reinforces and is reinforced by her sense that the conceptual problems are useless. Note that this is a toy model, leaving out the vast majority of cognitive elements needed to model her cognition. The arrows represent excitatory links between nodes. The nodes in our model correspond to cognitive elements which can be at various grain-sizes [2],[28]. Specifically, in this model in figure 2, the nodes are epistemological. Judy's annoyance in this model would be peripheral to her epistemology. However, in the next subsection, we argue how such an affect-free model is insufficient to model Judy's cognition.

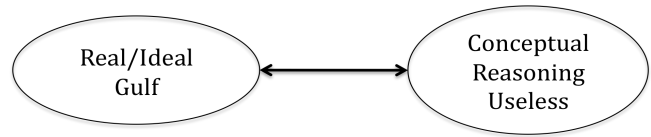

FIGURE 2
"TOY" COGNITIVE MODEL FOR JUDY'S EPISTEMOLOGICAL STANCES – NOT INCORPORATING AFFECT

*Judy's annoyance stabilizes her epistemological stance about the real/ideal gulf*

Two other episodes from the interview emphasize the need to incorporate affect into the cognitive toy model of Judy's reasoning.

*Disconfirmatory **Episode A**:* At the close of the interview, the interviewer asks Judy one of the conceptual homework questions from her course.

[50:06] Interviewer: Suppose you were comparing, um, three different types of waves in a circuit. So, you're comparing a sine wave, a square wave, and a triangle wave. And all three of them have the same RMS voltage. Which one has the highest peak voltage?

[50:36] Judy: {*sighs, smiles*} Um, triangle wave.

[50:39] Interviewer: OK. How do you know?

Judy responds by summarizing what she remembered of her TA's explanation, acknowledges that she can't remember it fully, and turns to an equation. She talks about how the answer has something to do with the integral, which is "related to" the area under the curve. The interviewer then presses her to think it through for herself. In response, she does some excellent informal reasoning:

[52:58] Interviewer: So, if you didn't have your book, but you know there was some relationship between the integral and the area. Is there a way that you could think through this?

[53:07] Judy: Um. {*6 second pause*} I guess—I mean I can try. So they have the same period like from, uh zero to T. So the integral would be like from zero to t. Um, and then they have some kind of wave form here. And they have a same answer. So. {*5 second pause*} I dunno. I mean, if they have the same area, um and, this part is like the same length, um, because like. This looks like wider {*spreads hands apart, past her shoulders*}—like the area looks wider {*spreads hands again*}. So there should be, like. If you squeeze it into {*presses palms toward each other*} this form, it should be like, going up {*forms a peak with her hands*}. So, it should be like this. So, apparently, like, they have a higher peak.

Judy here is thinking of the area under the curve as a kind of squeezable stuff; since the area under the three curves is the same, the amount of stuff is the same, and when you "squeeze" a rectangle (square wave) into a triangle (triangle wave) of the same base length, the triangle ends up with a "higher peak."

[54:42] Interviewer: But I'm interested in that you said "squeezing." Do you ever

Judy: {*Judy laughs*} no no no,

Interviewer: do you ever um, ever um think about the mathematics that you use in that way?

[54:53] Judy: No, I never think of that before. That's the way how the—my TA told me. Yeah, I mean it's I feel like it's not very formal {*smiles*}, but it's very useful.

Our transcript does not capture the satisfaction—evident in the video—she felt about her solution. Though Judy says of her solution, "it's not very formal"—it's the kind of qualitative reasoning she earlier labeled unproductive—in this moment, feeling good about her solution, she calls this kind of conceptual reasoning "very useful." The toy model in figure 2 doesn't predict such a reversal of her epistemological stance: if *annoyance* didn't mediate the link between the other two elements, then her positive experience should not have changed her view (ephemerally or otherwise) about the uselessness of conceptual reasoning.

*Disconfirmatory **Episode B***: At one point, not discussed above, while Judy was still thinking about her annoyance at the conceptual problems, the interviewer probes her views about *both* the real/ideal gulf *and* conceptual reasoning: "So do you think if you're analyzing a real-world circuit, it's important to know about the physical aspects of the circuit?" Judy responds, "Not very important." Crucially, her annoyance at the conceptual problems and the associated conceptual reasoning are still apparent during this part of the interview. The two-node model in figure 2 cannot explain this: a model in which she finds conceptual problems unproductive because they apply only to an idealized world does not explain why she finds conceptual reasoning unproductive even when used to analyze real-world circuits.

Of course, an affect-free model different from ours in Fig. 2 might be able to explain Judy's reasoning patterns throughout the interview. But given (i) the previous literature showing the role of affect on epistemological stances, and (ii) the in-your-face nature of the annoyance Judy displays towards conceptual problems during most of the interview, we have reason to take seriously models in which this affective state plays a role in stabilizing or destabilizing Judy's epistemological stances. More specifically, we argue that in her cognition during this interview, annoyance mediates the connection she draws between conceptual problems and the gulf between real/ideal worlds. Our proposed three-node toy model (figure 3) represents this: *annoyance* mediates *real/ideal gulf* and *conceptual-reasoning-useless* nodes without those epistemological nodes connected directly. This new model, incorporating an affective state (*annoyance*) can explain Judy's behaviors and reasoning:

- Her *annoyance* at conceptual questions triggers her stance towards conceptual reasoning as useless and the real/ideal gulf (Patterns#2 and #3),
- When she perceives herself as not engaged in conceptual reasoning but problem solving, her annoyance disappears and so does the idea of real/ideal gulf in those moments (even though the equations she is using encode idealizations) (Pattern #4).
- Suppressing *annoyance* can suppress the activation of "*conceptual reasoning useless*" leading to the reversal of her epistemological stance (Episode A), and
- Targeting Judy's epistemology (by focusing on *real* circuits) without addressing Judy's affect fails to suppress Judy's view that conceptual reasoning is "not very important" (Episode B).

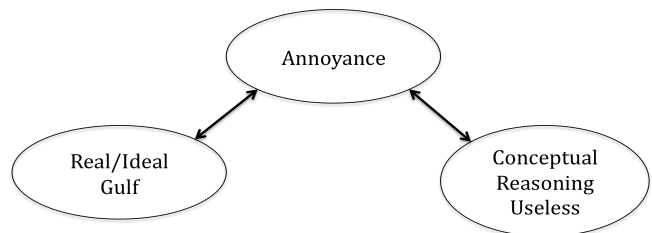

FIGURE 3
INCORPORATING AFFECT BETTER EXPLAINS THE CONTEXTUAL DYNAMICS OF JUDY'S EPISTEMOLOGICAL STANCES

## SUMMARY AND IMPLICATIONS

At first glance, Judy's patterns of thought (discussed above) seem attributable in part to a robust belief that ideal circuits (addressed by conceptual problems on the homework) are very different from real circuits. We argued, however, that this "belief" is not a robust part of Judy's reasoning across all contexts probed in the interview. Rather, an affective state—annoyance—is the bridge linking her views about the course's conceptual problems and her epistemological stance about a real/ideal gulf. Including a context-dependent affective state (annoyance) in our model enabled us to explain: (i) the minimal or nonexistent role her epistemological stance (real/ideal gulf) plays in her thinking when she discusses or performs mathematical problem-solving, and (ii) the disruption of her stance toward conceptual reasoning induced by an affectively positive experience. In this way, we illustrated how including affect in models of cognitive dynamics provides additional explanatory power.

This analysis has instructional implications. Given Judy's stated belief that the conceptual problems are annoying because of the real/ideal gulf, an obvious instructional strategy might be to directly address the real/ideal gulf by showing how idealizations approximate the real world, stressing the near-agreement of ideal predictions with real results. While we don't disagree with that strategy, we don't think it would be enough for students like Judy. Bringing about epistemological change for Judy (and students like her) might have more to do with setting her up to have affectively positive experiences, and less to do with eliciting, confronting, and replacing her problematic epistemological "beliefs." Ultimately, though limited to one student in one interview, our analysis illustrates the importance and feasibility of incorporating affect into fine-grained models of cognitive dynamics and how doing so can expand the toolbox of productive instructional responses.

## ACKNOWLEDGEMENTS

We thank "Judy" for her participation in the study. We thank the instructor of the Basic Circuits course for allowing us access to students, and incorporating suggested modifications into the curriculum. We thank Eric Kuo, Michael Hull, David Hammer, and members of the Physics Education Research Group for productive discussions. This work was supported in part by NSF EEC-0835880 and NSF DRL-0733613. The opinions in this manuscript are of the authors only.